\documentclass[12pt]{article}
\usepackage{graphicx}
\usepackage{bm}
\usepackage{amssymb,amsmath}
\begin{document}
\begin{titlepage}

\vspace{1cm}

\begin{center}
{\Large {\bf Coulomb effects in the spin-dependent\\contribution to the intra-beam scattering rate}}\\
\vspace{1.0cm}  V.M. Strakhovenko
\footnote { e-mail: v.m.strakhovenko@inp.nsk.su}\\
\vspace{0.5cm} {\it Budker Institute of Nuclear Physics, 630090,
Novosibirsk, Russia}
\end{center}
\vspace{2cm}
\noindent keywords: electron beam, internal scattering, polarization, Coulomb effects

\vspace{2.0cm}

\begin{abstract}

\noindent Coulomb effects in the intra-beam scattering are taken into account in a way providing correct description of the spin-dependent contribution to the beam loss rate. It allows one to calculate this rate for polarized $e^{\pm}$ beams at arbitrarily small values of the ratio $\delta \varepsilon/\varepsilon$, characterizing relative change of the electron energy in the laboratory system during scattering event.

\end{abstract}
\end{titlepage}
\newpage

\section{Introduction}
As is well known, elastic scattering of electrons in the beam moving in a storage ring leads to a beam loss (Touschek effect). The longitudinal momentum $p_{\parallel}$ of an electron is negligibly small in the beam rest system (RS) due to the small energy spread in the beam. However, two electrons subjected to scattering may get in RS such $p_{\parallel}^\prime$ that addresses in the laboratory system (LS) the relative energy deviation $\pm \delta \varepsilon/\varepsilon$ which is larger than the permissible one. Then both electrons leave the beam and can be registered by some counters. The rate of such events depends on the beam polarization since scattering cross section depends on the polarization of electrons. So, as was proposed in \cite{BaierHose}, the beam polarization can be measured by measuring that rate.  The resonant depolarization technique is developed in BINP for such measurements since 1970 (see e.g. \cite{BaierUFN}, \cite{Serdnyak76}). In this technique, a jump in counting rate of Touschek electrons or positrons is measured. It occurs when initially polarized beam is rapidly depolarized by applying the magnetic field with a frequency satisfying the spin resonance conditions. As the resonant frequency depends in a certain way on the beam energy, the latter can be measured alongside with the beam polarization. Such a technique is routinely used in BINP for the high accuracy absolute calibration of the beam energy (see e.g. \cite{SerSid77}, \cite{Blinov02}).

First spin effect calculations were carried out in \cite{BaierHose} assuming that the betatron motion is one-dimensional (flat beam) and the ratio $\delta \varepsilon/\varepsilon$ is very small. Both limitations were overcome in \cite{Serdnyak76}. However, some formulas in \cite{Serdnyak76} are not quite accurate. The scattering cross section used in both papers was obtained in the Born approximation. Accurate formulas for the scattering rate were derived in \cite{BKS78} within the same approximation. Coulomb effects were also considered in \cite{BKS78} but only for the flat beam. In the case of the electron-electron interaction these effects are important at very small velocities $v$ in the center-of-mass system (c.m.s.) when $\alpha/v\gtrsim 1$ ($\alpha$ is the fine-structure constant and $\hbar=c=1$). For such velocities the cross section is modified due to a corresponding change of the $e^+$ ($e^-$) flux near the origin (see, e.g. \cite{LL}). In particularly, the spin-dependent terms in the cross section are enhanced at $\alpha/v\gg 1$ for attraction and suppressed for repulsion (see discussion in \cite{MilSS}). In the intra-beam scattering we are just dealing with repulsion.

In the present paper, the scattering cross section is modified in such a way that at $v\ll 1$ it becomes the nonrelativistic one (see, e.g. \cite{LL}) taking the Coulomb effects into account.  The modified cross section passes into the expression obtained in the Born approximation at $\alpha/v\ll 1$, being thereby correct for any $v$. To get the rate of events with $\mid \delta \varepsilon\mid/\varepsilon$ in c.m.s. larger than some value, $\eta$, the differential cross section should be first integrated over the region where $\mid v_{\parallel}'\mid \geqslant \eta$. This condition can be satisfied only if $v \geqslant \eta$. Thus $v\ll 1$ may enter the problem for $\eta \ll 1$. However, use of the modified differential cross section provides one with values of the integrated cross section, $\sigma_\eta$, which are correct for any $\eta$.

\section{Results and discussion}
Let us consider a collision of two electrons of the beam, having in LS momenta ${\bm p}_{1,2}$, energies $\varepsilon_{1,2}$, and polarization vectors ${\bm\zeta}_{1,2}$. It is convenient to calculate the scattering cross section in their c.m.s. A transition to that system is performed by the Lorentz transform with velocity ${\bm V}=({\bm p}_1+{\bm p}_2)/(\varepsilon_1+\varepsilon_2)$. Conditions $\varepsilon/m \gg 1$ and ${\bm p}_\perp^2/\varepsilon^2\ll 1$, where $m$ is the electron mass, are well fulfilled in storage rings. Then we have in c.m.s., where ${\tilde{\bm p}}_1+{\tilde{\bm p}}_2=0 $,
$$ \tilde{p}_1^{\parallel}\simeq \tilde{\varepsilon}\,\frac{\varepsilon_1^2-\varepsilon_2^2}{ 4\varepsilon_1 \varepsilon_2} \sim \frac{1}{2}\tilde{\varepsilon}\,\left(\frac{\Delta\varepsilon_b }{\varepsilon_b}\right)\,,$$
where $\varepsilon_b $ is the mean beam energy. Perpendicular and parallel components of vectors are defined here with respect to the velocity  ${\bm V}$, which direction practically coincides with the beam axis. In our problem, $ \tilde{p}_1^\parallel$ can be neglected due to the smallness of the relative energy spread in the beam, $r=\Delta\varepsilon_b /\varepsilon_b$. More precisely, let $ \tilde{p'}_1^\parallel$ be the parallel component of the electron momentum in c.m.s. after scattering. Then its energy in LS becomes $\varepsilon'_1\simeq \varepsilon_b\,(1+\tilde{p'}_1^\parallel/\tilde{\varepsilon})$ and the relative  energy deviation $d= \delta \varepsilon_1/\varepsilon_1 \simeq \tilde{p'}_1^\parallel/\tilde{\varepsilon}$. We consider events with $\mid d \mid \geqslant \eta$ for $\eta$ providing a loss of particles. The latter means that $\eta \gg r$. So, the minimum value of $ \tilde{p'}_1^\parallel$ which enter the problem is much larger than the initial value of $ \tilde{p}_1^\parallel$. Self-consistently, the transverse momentum turns out to be much larger than the parallel one as $ \tilde{p}_1^\parallel /\tilde {p}_{1\perp} \sim r/v$ and $v \geqslant \eta $. In what follows we set $ \tilde{p}_1^\parallel =0$, so that the electron energy is $\tilde{\varepsilon}=\varepsilon_q\equiv\sqrt{{\bm q}^2+m^2}$, where ${\bm q}= ({\bm p}_{1\perp}-{\bm p}_{2\perp})/2={\bm{\tilde p}}_{1\perp}$.

We start from the well known expression for the invariant event number in a collision of two beams with spatial densities $n_{1,2}({\bm r})$ and momenta ${\bm p}_{1,2}$
\begin{equation}\label{invev}
dn= I\frac{n_1({\bm r})n_2({\bm r})}{\varepsilon_1\varepsilon_2}d\sigma d^3r dt\,,\quad I=\sqrt{(p_1 p_2)^2-m^4}\,.
\end{equation}
In our case $\varepsilon_1\simeq\varepsilon_2\simeq \varepsilon_b$, the invariant $I\simeq 2q \varepsilon_q$, $n_1({\bm r})=n_2({\bm r})\equiv n({\bm r})$ and $d\sigma$ is  the differential cross section for elastic $ee$ scattering. To get the total rate of producing Touschek pairs, we have to integrate $dn$ in Eq.(\ref{invev}) over ${\bm r}$, calculate $\sigma_ \eta$ by integrating $d\sigma$ over the region $\mid v_{\parallel}'\mid \geqslant \eta$, and average over momentum and polarization distribution in the beam. The result should be divided by two as we are dealing with identical particles. For Gaussian distribution over coordinates we have
\begin{equation}\label{intcor}
\int d^3r n^2({\bm r})=\frac{N^2}{V_b}\,,\quad V_b=8\pi^{3/2}\Delta_x \Delta_z \Delta_\parallel\,,
\end{equation}
where $N$ is the total number of particles in the beam, $ V_b$ is the beam volume, and $\Delta$s are
r.m.s. radial ($x$), vertical ($z$), and longitudinal ($\parallel$) sizes of the beam, respectively. To average over momenta, the integral $\int d^3p_1 d^3p_2 f({\bm p}_1) f({\bm p}_2)q \varepsilon_q \sigma_ \eta$ should be taken. Here $ f({\bm p})$ represents the momentum distribution in the beam.  Since ${\bm p}_{1,2}$ enter $\sigma_ \eta$ and the invariant $I$ only in the combination ${\bm q}= ({\bm p}_{1\perp}-{\bm p}_{2\perp})/2$, the integrals over $ p_{1,2}^\parallel$ and $({\bm p}_{1\perp}+{\bm p}_{2\perp})$ can be easily taken. After that the integral passes into $\int d^2q \sigma_ \eta ({\bm q})q \varepsilon_q F({\bm q})$. Using the Gaussian type of $ f({\bm p})$ with r.m.s. parameters $\delta_x$, $\delta_z$, and $ \delta_\parallel$, we obtain $F({\bm q})=(\pi \delta_x \delta_z)^{-1}\exp(-q_x^2/\delta_x^2-q_z^2/\delta_z^2)$ as the distribution over ${\bm q}$. So the rate in LS, $\nu(\eta)$, reads
\begin{equation}\label{rategen}
\nu(\eta)\equiv \frac{dn}{dt}=\frac{N^2}{\pi \delta_x \delta_z V_b \varepsilon_b^2}\int d^2q \sigma_ \eta ({\bm q})q \varepsilon_q\exp(-q_x^2/\delta_x^2-q_z^2/\delta_z^2)
\end{equation}
Now we pass to the differential cross section for elastic scattering of polarized electrons summed up over final spin states. Let us define in c.m.s. the coordinate system with basis vectors ${\bm e}_3={\bm q}/q$, ${\bm e}_2={\bm V}/V$, and ${\bm e}_1=[{\bm e}_2 \times {\bm e}_3]$. The scattering angles $\vartheta$ and $\phi$ are defined in such a way that $\tilde {{\bm p}}'/q={\bm e}_3\cos\vartheta+({\bm e}_1\cos\phi+{\bm e}_2\sin\phi)\sin\vartheta $.

In that notation the cross section reads
\begin{equation}\label{Sigmadif}
\frac{d\sigma}{d\Omega'}=\frac{\alpha^2}{2\varepsilon_q^2}(F_1+F_2+F_3+F_4)\,,
\end{equation}
where the functions $F_i$ are
\begin{eqnarray}\label{Func14}
F_1&=&2m^4\Bigl [\frac{1}{t^2}+\frac{1}{u^2}+\frac{1+{\bm\zeta}_1\cdot{\bm\zeta}_2}{tu}\cos\left(\frac{\alpha}{2v}\ln\frac{t}{u} \right)\Bigr] \,,\nonumber\\
F_2&=&8q^2\varepsilon_q^2\Bigl (\frac{1}{t^2}+\frac{1}{u^2}\Bigr)+\frac{1}{2}(1+{\bm\zeta}_1\cdot{\bm\zeta}_2)\left(1-\frac{16m^2q^2}{tu}\right)
+\left(\frac{m^2}{q^2}-1\right){\bm e}_3\cdot{\bm\zeta}_1{\bm e}_3\cdot{\bm\zeta}_2\,,\nonumber\\
F_3&=&\Bigl [\frac{4q^2(4q^2+m^2)}{ut}-1\Bigr]\Bigl [{\bm e}_2\cdot{\bm\zeta}_1{\bm e}_2\cdot{\bm\zeta}_2-\frac{m}{q}{\bm e}_1\cdot[{\bm \zeta}_1 \times {\bm \zeta}_2]\Bigr]\,,\\
F_4&=&\Bigl (\frac{4q^2m^2v^2}{ut}-1\Bigr)\Bigg \{\sin^2\phi\Bigl [\frac{m^2}{q^2}{\bm e}_2\cdot{\bm\zeta}_1{\bm e}_2\cdot{\bm\zeta}_2-{\bm e}_3\cdot{\bm\zeta}_1{\bm e}_3\cdot{\bm\zeta}_2+\frac{m}{q}{\bm e}_1\cdot[{\bm \zeta}_1 \times {\bm \zeta}_2]\Bigr]\nonumber\\
&+&\frac{\cos^2\phi}{v^2}{\bm e}_1\cdot{\bm\zeta}_1{\bm e}_1\cdot{\bm\zeta}_2\Bigg \}\,,\nonumber
\end{eqnarray}
where $t=-2q^2(1-\cos\vartheta)$ and $u=-2q^2(1+\cos\vartheta)$ are conventional kinematic variables. The expression (\ref{Sigmadif}) is derived in the following way. We start from the result obtained in \cite{FordM} in the Born approximation and unfold that in c.m.s. As was noted in \cite{BKS78}, while unfolding in c.m.s. the invariants containing 4-vector of spin, one should remember that the unit spin vector ${\bm\zeta}$ is not  Lorentz invariant. If in some reference frame an electron has the momentum ${\bm p}$, the energy $\varepsilon$, and the spin vector ${\bm\zeta}$, then in the reference frame moving with the velocity ${\bm V}$ the vector  ${\bm\zeta}'$ reads
\begin{equation}\label{zeta}
{\bm\zeta}'={\bm\zeta}+{\bm\nu}({\bm\zeta}\cdot[{\bm a} \times {\bm b}])-[{\bm \nu} \times {\bm b}]{\bm\zeta}\cdot{\bm a}\,,
\end{equation}
where
\begin{equation}\label{dzetapr}
\bm\nu={\bm V}/V,\quad{\bm a}={\bm p}\frac{\gamma_V-1}{\varepsilon+m}-{\bm V}\gamma_V,\quad{\bm b}= \frac{[{\bm p} \times {\bm\nu}]}{\varepsilon'+m},\quad\gamma_V=\frac{1}{\sqrt{1-V^2}}\,.
\end{equation}
As a next step, the terms giving the nonrelativistic limit (for ${\bm v}={\bm q}/\varepsilon_q\ll 1$) are separated up. In Eq.(\ref{Sigmadif}) such terms are combined in the function $F_1$ which is proportional to $v^{-4}$. The rest functions in Eq.(\ref{Sigmadif}) contain terms having at least excess factor of $v^2$ as compared with $F_1$. They represent in this limit the relativistic corrections. In the Born approximation, we have for $F_1^B$
\begin{equation}\label{f1Born}
F_1^B=2m^4\Bigl [\frac{1}{t^2}+\frac{1}{u^2}+\frac{1+{\bm\zeta}_1\cdot{\bm\zeta}_2}{tu}\Bigr]
\end{equation}
The last term in Eq.(\ref{f1Born})($\propto) 1/tu$) addresses the exchange interaction. It originates from interference of two Coulomb amplitudes addressing scattering at $(\vartheta,\phi)$ and  $(\pi-\vartheta, \pi+\phi)$. A phase of the Coulomb amplitude is proportional to $\alpha$ and is ignored in the Born approximation. Taking that phase and thereby the Coulomb corrections into account results in the appearance of the factor $\cos\left(\frac{\alpha}{2v}\ln\frac{t}{u} \right)$ by the exchange term in Eq.(\ref{f1Born}). In that way we pass from $F_1^B$ in Eq.(\ref{f1Born}) to $F_1$  in Eq.(\ref{Func14}). That modification provides the correct nonrelativistic limit of the cross section (\ref{Sigmadif}). Strictly speaking, the rest functions in Eq.(\ref{Sigmadif}) are also changed by the Coulomb corrections. However, since the latter become apparent only in the nonrelativistic limit, where the terms given by functions $F_{2,3,4}$ are small as compared with corresponding terms in $F_1$, in our problem we can use expressions for $F_{2,3,4}$ obtained in the Born approximation. Let us emphasize that sometimes Coulomb effects  might be important also in terms representing relativistic corrections. For example, to estimate the beam depolarization during the intra-beam scattering one should calculate a probability of spin flip transitions. The spin dependence of the nonrelativistic cross section is completely due to the exchange interaction. In full agreement with the sense of that interaction it allows only the simultaneous spin flip of both electrons, at that the initial spin directions should be opposite. Such transitions do not change the beam polarization. Therefore, relativistic corrections, addressing direct spin-spin and spin-orbit coupling, should be considered. As is shown in \cite{MilSS}, Coulomb effects  may be very important in that case.

\begin{figure}[h]
\centering
\includegraphics[width=0.8\textwidth]
{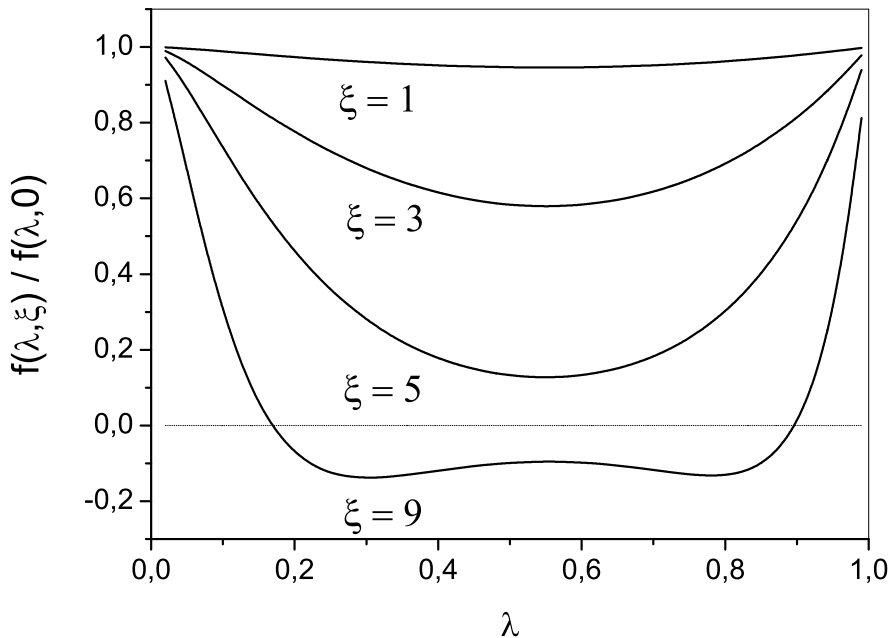}
\caption{Ratio of the function $f(\lambda,\xi)$
to its limit value  $f(\lambda,0)=\ln 1/\lambda $ . } \label{Fig:flambda}
\end{figure}

Passing to the calculation of $\sigma_ \eta$, we integrate the differential cross section (\ref{Sigmadif}) over the region $\mid v_{\parallel}'\mid \geqslant \eta$. In terms of the angles $\vartheta$ and $\phi$, the latter condition reads $\sin^2\phi\sin^2\vartheta\geqslant \lambda^2$, where $\lambda=\eta/v\leqslant 1$. The only change in $\sigma_ \eta$ as compared to $\sigma_ \eta^B$  comes from the exchange term in $F_1$. Namely, the integral $\int d\Omega'(tu)^{-1}\Theta(\sin^2\phi\sin^2\vartheta - \lambda^2)=\frac{\pi}{q^4}\ln \frac{1}{\lambda}$ passes into $\int d\Omega'(tu)^{-1}\cos\left(\frac{\alpha}{2v}\ln\frac{t}{u} \right)\Theta(\sin^2\phi\sin^2\vartheta - \lambda^2)=\frac{\pi}{q^4}f(\lambda,\xi)$. Here $\Theta(x)$ is the step-function and
\begin{equation}\label{flambda}
f(\lambda,\xi)=\frac{2}{\pi}\int\limits_0^{\operatorname{Arcosh}\frac{1}{\lambda}}dx \cos(\lambda\xi x)\arccos(\lambda\cosh x)\,\equiv\,\frac{2}{\pi\lambda\xi}\int\limits_0^{\arccos\lambda}dx \sin\left [\lambda\xi \operatorname{Arcosh} \left(\frac{\cos x}{\lambda}\right)\right]\,,
\end{equation}
where $\xi=\alpha/\eta$ and $\operatorname{Arcosh}z=\ln(z+\sqrt{z^2-1})$. Shown in Fig.\ref{Fig:flambda} is
the ratio of the function $f(\lambda,\xi)$ to its limit in the Born approximation, $f(\lambda,0)=\ln 1/\lambda $.
As is seen on Fig.\ref{Fig:flambda}, at $\xi=1$ the function $f(\lambda,\xi)$ practically coincide as yet with its limit at $\xi\rightarrow 0$. At $\xi\gg 1$ the function $f(\lambda,\xi)$ is rapidly oscillating and the corresponding integral over $\lambda$ ($q$) in (\ref{rategen}) falls off as $\xi^{-2}$.

Let the beam polarization be directed along the unit vector ${\bm s}$. We choose that direction as the quantization axis while performing averaging over electron polarizations. Then we obtain for the combination $\zeta_1^i \zeta_2^j$ which enter the cross section: $<\zeta_1^i \zeta_2^j> = P^2 s^i s^j$, where $P$ is the beam polarization degree. By definition, $P= (N_\uparrow-N_\downarrow)/N$, where $N_\uparrow$ and $N_\downarrow$ are numbers of electrons with spin projections onto the quantization axis $\pm 1/2$, respectively. The final form of $\sigma_ \eta$ reads $$\sigma_\eta({\bm q})=\frac{\pi\alpha^2}{q^4\varepsilon_q^2}G({\bm q})\Theta(1-\lambda) \,,$$ where
\begin{equation}\label{funG}
G({\bm q})=A_0+q^2 A_1+P^2\left[ q^2s_\parallel^2 A_\parallel+q^2(s_x^2+s_z^2)A_2+(q_x^2-q_z^2) (s_z^2-s_x^2)A_3\right]\,.
\end{equation}
Here
\begin{eqnarray}\label{Afunc}
A_0&=&(2q^2+m^2)^2\left(\frac{1}{\lambda^2}-1\right)-m^4(1+P^2)f(\lambda,\xi)\,,\nonumber\\
A_1&=&q^2(1-\lambda)-4m^2\ln\frac{1}{\lambda}\,,\\
A_\parallel&=&(1+v^2)\left[(4\varepsilon_q^2-m^2)\ln\frac{1}{\lambda}-\frac{m^2}{2}(1-\lambda^2)\right]-2q^2 (1-\lambda)+A_1\,,\nonumber\\
A_2&=&m^2\left(\frac{m^2}{2\varepsilon_q^2}-4\right)\ln\frac{1}{\lambda}+\frac{(2q^2+m^2)^2}{4\varepsilon_q^2} (1-\lambda^2)\,,\nonumber\\
A_3&=&\frac{m^2}{2}(1+v^2)\left[\ln\frac{1}{\lambda}+\frac{1}{2}(1-\lambda^2)\right]-2m^2(1-\lambda)\,.\nonumber
\end{eqnarray}
Remember that $\lambda=\eta/v$, $v=q/\varepsilon_q$, and $f(\lambda,\xi)$ is defined by Eq.(\ref{flambda}). Substituting the obtained expression for $\sigma_ \eta$ into Eq.(\ref{rategen}), we have for the rate
\begin{equation}\label{ratefin}
\nu(\eta)=\frac{\alpha^2 N^2}{\varepsilon_b^2 V_b\delta_x \delta_z }\int \frac{ d^2q}{q^3 \varepsilon_q}G({\bm q})\exp(-q_x^2/\delta_x^2-q_z^2/\delta_z^2)\Theta(1-\lambda)\,.
\end{equation}
If we substitute $\ln1/\lambda$ for $f(\lambda,\xi)$ in $A_0$, Eq.(\ref{ratefin}) passes into Eq.(5) of \cite{BKS78}. The expression (\ref{ratefin}) represents the rate of producing in the beam a pair of electrons with relative energy deviation $\pm \delta \varepsilon/\varepsilon_b$ which module is larger than $\eta$. It has a local character as the beam parameters and the polarization direction ${\bm s}$ may vary along the orbit. An interval of $\eta$, which is important in measurements of the rate, is determined by a size and a position of a counter. Corresponding  $\eta$s may be small or not. If we apply Eq.(\ref{ratefin}) to estimate the beam lifetime, we shell deal with small $\eta\ll 1$. For any interrelation between $\delta_x$, $\delta_z$, and $\eta m$  the numerical integration in Eq.(\ref{ratefin}) is very simple. It is straightforward for the flat or round beam and for any $\eta$. Nevertheless, let us consider separately small $\eta\ll 1$. All terms in $G({\bm q})$ except $A_0$ can be neglected in that case. Going from the integration over $q$ to the integration over $\lambda$, we obtain for $\eta\ll 1$
\begin{equation}\label{smallet}
\nu(\eta)=\frac{2\pi r_e^2 N^2 m^2}{\gamma_b^2 V_b\delta_x \delta_z \eta^2}\int\limits_\eta^1 dz\left\{\left(\frac{1+z^2}{1-z^2}\right)^2 g\left(\frac{z^2}{1-z^2}\right)-\eta[1+(1+P^2)f(z,\xi)]g\left(\frac{\eta^2}{z^2-\eta^2}\right)\right \},
\end{equation}
where $\gamma_b=\varepsilon_b/m$ and
\begin{equation}\label{gbeta}
g(Y)=\exp(-Y\beta_+^2)I_0(Y\beta_-^2),\quad \beta_\pm^2=\frac{m^2}{2}\left(\frac{1}{\delta_z^2}\pm\frac{1}{\delta_x^2}\right).
\end{equation}
Here $I_0(x)$ is the modified Bessel function. Note that the rate (\ref{smallet}) is independent of the beam polarization direction ${\bm s}$.

\begin{figure}[h]
\centering
\includegraphics[width=0.8\textwidth]
{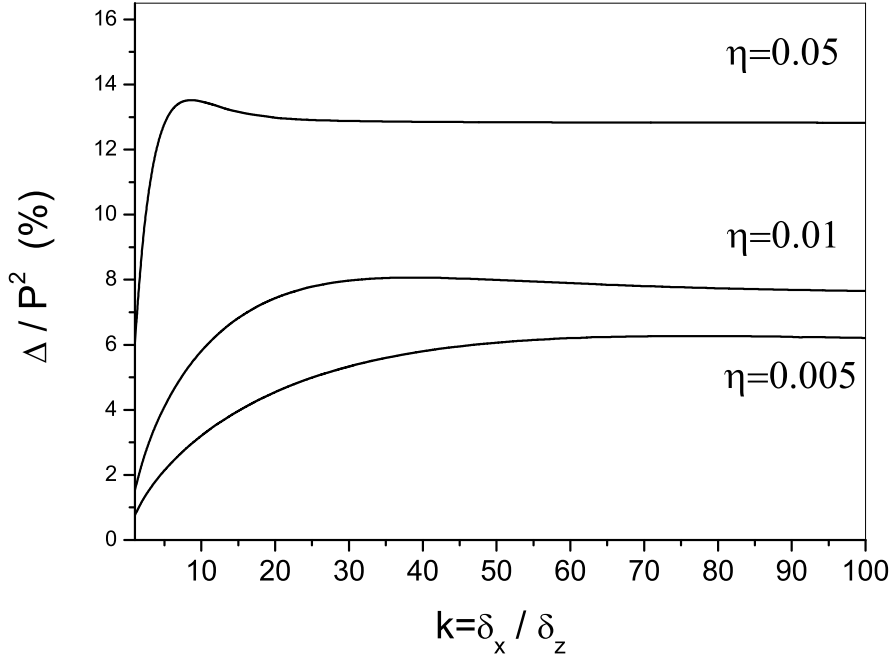}
\caption{Dependence of maximal jump $\Delta/P^2$ on $k=\delta_x/\delta_z$
at $\delta_x=0.5m $. } \label{Fig:jump}
\end{figure}

Let Touschek electrons with relative energy deviation larger than $\eta$ are registered by some counter. A jump in the counting rate is observed when initially polarized beam is depolarized. The jump $\Delta$ is usually defined as $\Delta=1-\nu(\eta,P^2)/\nu(\eta,0)$. In Fig.\ref{Fig:jump} the maximal possible jump $\Delta/P^2$ is shown as a function of $k=\delta_x/\delta_z$. It was calculated using Eq.(\ref{smallet}) for several small values of $\eta$ at $\delta_x=0.5m $. From Fig.\ref{Fig:jump}, the jump increases with $\eta$ and the regime of the flat 
beam is achieved earlier for larger $\eta$. From Eq.(\ref{ratefin}) we conclude that in our problem such regime is realized at arbitrary  $\eta$ for $\delta_z \ll \eta m$ and $\delta_x \gtrsim \eta m$. If both $\delta_z$ and $\delta_x$ are much smaller than $\eta m$, the rate (\ref{ratefin}) is strongly suppressed being proportional to $\exp (-2\eta^2\beta_+^2)$. In the case of the flat beam and small $\eta$ we obtain from  Eq.(\ref{smallet})
\begin{equation}\label{flatb}
\nu(\eta)=\frac{2\sqrt{\pi} r_e^2 N^2 m}{\gamma_b^2 V_b\delta_x \eta^2}\left\{\ln\frac{2}{\eta}-\frac{3}{2}-\frac{c(\xi)}{4}(1+P^2)+B\left(\frac{m}{\delta_x} \right)\right\}\,,
\end{equation}
where
\begin{eqnarray}\label{bafunc}
B(z)&=&\sqrt{\pi}\left\{\left(\frac{2}{z}+z\right)[1-\Phi(z)]\exp(z^2)-\int\limits_0^z dx[1-\Phi(x)]\exp(x^2) \right\}\,,\nonumber\\
\Phi(z)&=&\frac{2}{\sqrt{\pi}}\int\limits_0^z dx\exp(-x^2) ,\qquad c(\xi)=\frac{4}{\xi}\int\limits_0^1 dx J_1\left(x\xi\operatorname{Arcosh}\frac{1}{x}\right)\,,
\end{eqnarray}
here $J_1(u)$ is the Bessel function. At $\xi\rightarrow 0$ the function $c(\xi)\rightarrow 1$ and it is suppressed at $\xi\gg 1$, where $c(\xi)\simeq 4/(\xi^2 \ln 2\xi)$. Such power suppression at $\xi\gg 1$  is an ordinary Coulomb effect for repulsion.

To conclude, the cross section of intra-beam scattering is modified to take into account Coulomb effects. It allows one to calculate the beam loss rate for polarized $e^{\pm}$ beams at arbitrary values of $\eta$.

\section{ACKNOWLEDGMENTS}
 This work was supported in part by the grant 09-02-00024 of the Russian Foundation for Basic Research.

\end{document}